\documentclass[sigconf]{acmart}

\AtBeginDocument{%
  \providecommand\BibTeX{{%
    \normalfont B\kern-0.5em{\scshape i\kern-0.25em b}\kern-0.8em\TeX}}}

\copyrightyear{2020}
\acmYear{2020}
\setcopyright{acmlicensed}
\acmConference[HRI '20]{Proceedings of the 2020 ACM/IEEE International Conference on Human-Robot Interaction}{March 23--26, 2020}{Cambridge, United Kingdom}
\acmBooktitle{Proceedings of the 2020 ACM/IEEE International Conference on Human-Robot Interaction (HRI '20), March 23--26, 2020, Cambridge, United Kingdom}
\acmPrice{15.00}
\acmDOI{10.1145/XXXXXX.XXXXXX}
\acmISBN{978-1-XXX-XXX-X/XX/XX}

\usepackage{verbatim}
\usepackage{amsmath}
\usepackage{tcolorbox}

\usepackage{booktabs} 
\usepackage{enumitem}
\usepackage{makecell}
\usepackage{textcomp}
\usepackage{amsmath}
\usepackage{algorithm}
\usepackage[noend]{algpseudocode}
\usepackage{pdfpages}
\usepackage{graphicx}
\usepackage{amssymb}
\usepackage{mathrsfs}
\usepackage{xcolor}
\usepackage{epsfig}
\usepackage{graphicx}

\author{David A. Robb$^1$$^*$, Muneeb I. Ahmad$^1$, Carlo Tiseo$^2$, Simona Aracri$^2$, Alistair C. McConnell$^2$,}

\author{Vincent Page$^3$, Christian Dondrup$^1$, Francisco J. Chiyah Garcia$^1$, Hai-Nguyen Nguyen$^4$,}
\author{ \`Eric Pairet$^{1,}$$^2$, Paola Ard\'on Ram\'irez$^{1,}$$^2$, Tushar Semwal$^2$, Hazel M. Taylor$^1$, Lindsay J. Wilson$^1$, David Lane$^1$, Helen Hastie$^1$, Katrin Lohan$^{1,}$$^5$}

\thanks{$^*$David Robb is the corresponding author, d.a.robb@hw.ac.uk}

\affiliation{\institution{$^1$Heriot-Watt University, $^2$University of Edinburgh,\\
$^3$University of Liverpool, $^4$Imperial College, $^5$NTB University of Applied Sciences in Technology, Buchs, CH}}

\begin{document}

\fancyhead{}

\title[Robots in the Danger Zone: Exploring Public Perception through Engagement]{Robots in the Danger Zone:\\ Exploring Public Perception through Engagement}

\renewcommand{\shortauthors}{D.A.Robb, et al.}

\begin{abstract}
Public perceptions of Robotics and Artificial Intelligence (RAI) are important in the acceptance, uptake, government regulation and research funding of this technology. Recent research has shown that the public's understanding of RAI can be negative or inaccurate. We believe effective public engagement can help ensure that public opinion is better informed. In this paper, we describe our first iteration of a high throughput in-person public engagement activity. We describe the use of a light touch quiz-format survey instrument to integrate in-the-wild research participation into the engagement, allowing us to probe both the effectiveness of our engagement strategy, and public perceptions of the future roles of robots and humans working in dangerous settings, such as in the off-shore energy sector. We critique our methods and share interesting results into generational differences within the public's view of the future of Robotics and AI in hazardous environments. These findings include that older peoples' views about the future of robots in hazardous environments were not swayed by exposure to our exhibit, while the views of younger people were affected by our exhibit, leading us to consider carefully in future how to more effectively engage with and inform older people.
\end{abstract}




\maketitle
\vspace{-3mm}
\begingroup
\fontsize{8pt}{8pt}\selectfont
\textbf{ACM Reference Format:}\\
David A. Robb, Muneeb I. Ahmad, Carlo Tiseo, Simona Aracri, Alistair C. McConnell, Vincent Page, Christian Dondrup, Francisco J. Chiyah Garcia, Hai-Nguyen Nguyen, \`Eric Pairet, Paola Ard\'on Ram\'irez, Tushar Semwal, Hazel M. Taylor, Lindsay J. Wilson, David Lane, Helen Hastie, Katrin Lohan. 2020. Robots in the Danger Zone: Exploring Public Perception through Engagement. In \textit{Proceedings of the 2020 ACM/IEEE International Conference on Human-Robot Interaction (HRI’20), March 23–26, 2020, Cambridge, UK.} ACM, New York, NY, USA, 10 pages. https://doi.org/10.1145/XXXXXX.XXXXXX
\endgroup


\section{Introduction}

Public engagement is widely appreciated and is regarded as essential for researchers and scientists for the wider dissemination of their work. It helps to establish public perception towards science and technology and is particularly important in the field of Robotics, Artificial Intelligence (RAI) and Human-Robot Interaction (HRI). More specifically, the field of HRI can make an important contribution to information transfer to the public.  
With the growing negative public perception in the field of RAI, more engagement building positive public perception is needed. A survey conducted in 27 European countries highlighted negative attitudes towards robots between 2012 and 2017. Notably, the concept of robots assisting in the work place received the strongest negativity \cite{gnambs2019robots}. One of the reasons behind these negative perceptions is increased media coverage and public discourse about robots replacing humans in several areas. A recent study analyzing the framing of Artificial Intelligence (AI) in American newspapers showed that, while the benefits of AI were highlighted more frequently in newspapers, AI's adverse traits were emphasized \cite{chuan2019framing}. 

\begin{figure}
\centering
  \includegraphics[width=1.0\columnwidth]{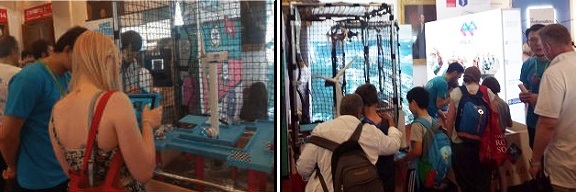}
  \caption{Crowds interacting with our exhibit.}~\label{fig:figureCrowds}
\end{figure}

We understand that this negative perception is also due to the misconceptions or limited knowledge of the public. Certainly, their knowledge is driven by print and digital media. A recent survey of 1078 participants conducted in the UK investigated awareness of RAI. The results showed that 85\% claimed to have heard of AI. However, it highlighted that most participants had a distorted understanding of AI \cite{cave2019scary}. The rationale for this limited knowledge lies in the complexity of forming public opinion, particularly in the cognitive miser theory \cite{orbell1991cognitive}. The theory suggests that people form opinions and attitudes with incomplete information using cognitive shortcuts or heuristics mostly gleaned from mass media \cite{scheufele2005public,orbell1991cognitive}. However, research shows that negative perceptions can be reduced by giving exposure or experience with the technology. For instance, in the education domain, a study conducted with teachers from five EU countries to understand their views on the use of robots in education showed that teachers found robots to be a disruptive technology \cite{serholt2014teachers}. Although, other researchers highlighted that teachers' views were derived from limited understanding or no experience of robots \cite{westlund2015interplay,ahmad2016understanding}. Importantly, they conducted studies and showed a transformation of perception after limited and long-term exposure to a robot in the classroom \cite{ahmad2016children,ahmad2016understanding,westlund2015interplay}.

We, therefore, believe that public engagement is important in building public perception and knowledge of AI and robots and it can play an important role in reducing negative perceptions. Consequently, we, a team of researchers from five universities took part in the prestigious Royal Society Summer Science Exhibition to disseminate information on RAI in hazardous environments. 


In this paper, we describe our approach to robotics engagement designed for a high throughput public event. This approach allows both engagement with the public and the conducting of research in the context of an event with more than 12,000 visitors over seven days (Figure \ref{fig:figureCrowds}). Our engagement activities were interactive and our research methods had to be light touch and seen as part of visiting the exhibit. The Royal Society event, at which we deployed our first iteration of this approach, had many exhibits and thus visitors typically had only a short time to engage at each. Our challenge was to entertain and inform all visitors who engaged at our RAI exhibit and expose as many as possible to research. The engagement challenge was met by designing a highly interactive yet safe exhibit allowing visitors of all ages to engage with researchers and get hands-on control of a robot to complete a challenge. It is important to note that the aim of our engagement was not to overtly influence opinion but to inform and thus indirectly influence public views of RAI. Our research aim was to explore and investigate whether knowledge and opinions on the usage of robots in hazardous environments differed between participants exposed and not exposed to the exhibit.  The research objective was met by devising a research protocol and instrument, which became part of the engagement itself for those who took part. The protocol was brief, lightweight, and applied the high ethical standards essential for involving members of the public including children \cite{Hay2006}.

\textbf{The contributions of this paper are summarized as:}
\begin{itemize}
\item Exposition of the methods used in a high throughput RAI public engagement activity.
\item The design of an engaging light touch survey instrument integrated with the engagement activity to probe public knowledge and opinions about the future of robotics and AI in extreme environments.
\item Results suggesting there are some generational differences in public views in this area.
\item Discussion of the results and of the effectiveness of the methods used for this mass in-person RAI public engagement activity and study.

\end{itemize}
\section{Background}
\subsection{Perception of Robotics and AI}

The perception of technology can have two broad aspects: 1) Functionality, and 2) Safety. A clear picture of both is required for a technology to be used safely. A misleading picture may result in catastrophes, such as recent airline crashes due to malfunctioning computer controlled systems \cite{vpref1,vpref2,vpref3,vpref4,vpref5}. 
Most people are not equipped with the training or prior knowledge to understand a system's abilities and safety as perhaps an engineer would. Their perceptions are often affected by high profile accidents and nefarious misuse delivered through the news media, films, TV programs and literature. 
Public knowledge about AI safety is shaped by comentary such as Azimov's Laws \cite{vpref6}, high profile books \cite{vpref7,vpref8}, and the open letter by popular scientists and engineers \cite{vpref9}.

The perception of RAI systems and technology is important, as it will determine first, whether the technology is used \cite{brougham2018smart,Peronard2013}, and second, whether it is used safely \cite{Weitschat2017,virk2014iso,winfield2018ethical}. If a negative view of RAI exists, there will be resistance to their use, adoption, and funding \cite{sundar2016hollywood}. 
This could then impact the possible safety benefits and increases in productivity and/or living standards that the RAI systems could provide. 
An important point to make here is that a blindly positive adoption of RAI is not always a good thing. It is argued that a perception mismatch, either an overly negative or positive view, can have negative impacts \cite{Hastie2017trusttriggers, Hastie2018, Lier2014}. While the focus of our engagement activity described in this paper is robotics in extreme environments, which has more clear-cut safety benefits to human well-being compared to RAI in general, we did still employ this balanced approach.
In addition, the climate of public opinion can influence research funding and regulatory policy \cite{Siegrist2000}. Good public engagement, presenting a balanced view, can help ensure that such influence is at least well-informed. 

It is challenging to inform public attitudes and beliefs. To understand some of the complexities, we reflect on two theoretical models from social psychology. First, the Cognitive Miser Model \cite{orbell1991cognitive}. This theorizes that people will only collect the minimum facts they think are needed and make decisions with limited information often relying on cognitive shortcuts such as attitudes and beliefs. Secondly, the Scientific Literacy Model \cite{bauer1993mapping}. This suggests that only those keenly interested in a technology seek more facts before taking an informed decision. Our activities in this paper were designed to provide relevant information in an engaging way for those activating the Scientific Literacy Model attending the exhibition. 

\subsection{Types of Robotics Engagement Activities}

A recent review on the usage of social robots in public spaces shows that they have primarily been used in the education and healthcare sector and also indicated their use in the shopping malls \cite{mubin2018social,serholt2013emote,foster2016mummer}.
Other projects include a semi-autonomous Heart Robot puppet \cite{rocks2009heart} and Zora robot for elderly care \cite{tuisku2019robots}. Most of the public engagements with these projects have been highlighting their usages in the various social domains such as care-homes, museums and similar others. The dissemination medium of public engagement with these robots has been mainly through social networking platforms, such as Twitter \cite{mubin2016naorobot}. The above highlights that, to the best of our knowledge, there has been limited systematic work published on the public engagement experience in the HRI community, particularly towards understanding the value of the engagement with robots in terms of shaping public opinions and augmenting their general knowledge in the area. Consequently, this demonstrates the need and value of the work presented in this paper as we aim to learn about the impact of a robotic engagement in shaping both the public's opinions and knowledge. 
\vspace{-2mm}
\section{The Engagement}
\subsection{Context and Aims}
The engagement took place at the prestigious annual week-long Royal Society Summer Science Exhibition, whose aim is to expose the public to the state-of-the-art research being done by universities. 
The exhibition attracted 12,653 visitors this year (2019).


The specific aim of the exhibit was to demonstrate how robots can operate both autonomously and also through remote control in hazardous environments typically unsafe for humans. A secondary aim was promoting STEM careers. To achieve these aims, we enabled visitors to engage with researchers and participate by completing a challenge through controlling the robots (details in Sections 3.3 and 3.4).




\vspace{-2mm}
\subsection{Design and Build}
 
A modular approach was adopted to allow for the following: a) the necessary portability of the exhibit; b) the accessibility and viewing angles for the visitors; c) the planned range of activities and engagement; and d) a level of flexibility e.g. ifa  module broke or proved more popular than the others. A fabrication company with experience in creating interactive exhibits was brought on board to build the exhibit. 

\textbf{There were three modules in the exhibit: 1) Wind Turbine, 2) Offshore Energy Platform, and 3) Industrial Maze.}

\begin{figure}
\centering
  \includegraphics[width=0.75\columnwidth]{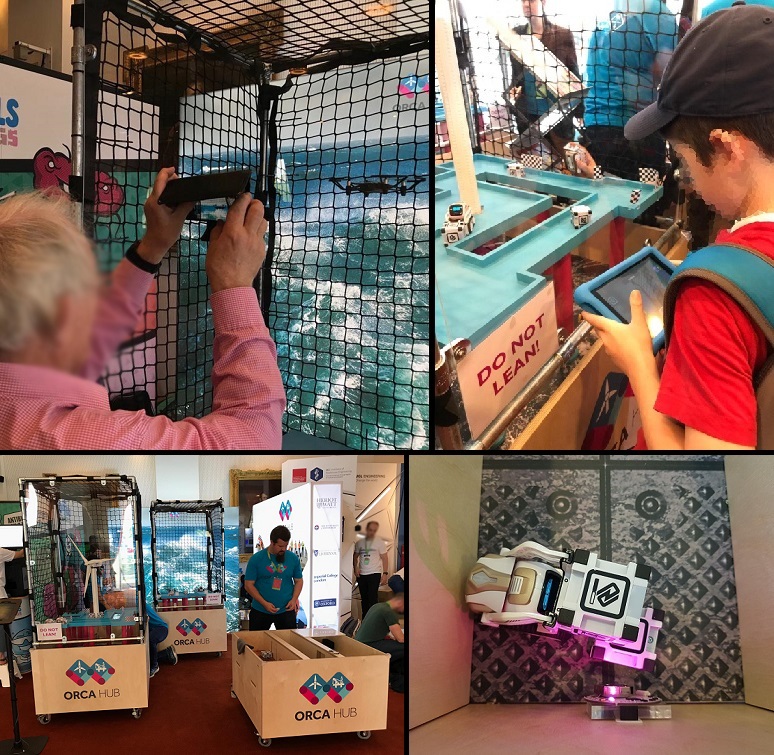}
  \caption{The three modules of the exhibit. Top Left: The Offshore Energy Platform module with a participant photographing a drone. Top Right: A child in front of the Wind Turbine module operating a Cozmo robot and attempting to return the tools (cube) to the correct location. Bottom Right:  A close up from above looking down into the Industrial Maze module, as a Cozmo robot performs a stacking maneuver. Bottom Left: All three modules with the Industrial Maze in the foreground and the Offshore Energy Platform in the background.  }~\label{fig:figureEngage1}
\end{figure}

\begin{figure}
\centering
  \includegraphics[width=0.8\columnwidth]{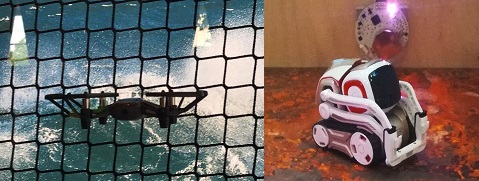}
  \caption{Left: One of the drones (approx 12 cm in diameter), in its netted enclosure. Right: A Cozmo (foreground) with a circular Limpet~\cite{sayed2018limpet} sensor attached to the wall behind.}~\label{fig:robotscombo}
\end{figure}

Each would present visitors with a different experience and allow them to consider the hazards of offshore environments \cite{hastie2018orcahub}. The Wind Turbine and Offshore Energy Platform modules were enclosed and would feature small robots (Cozmo\footnote{Made by Anki. Skageby\cite{Skageby2018} describes the background to its development.}, Limpet sensors~\footnote{A 6.5 cm diameter ROS enabled multi-sensing low-cost device, which can communicate with other ROS robots using wifi or LoRa\cite{sayed2018limpet}.} and Drones~\footnote{We used Tello EDU dones  https://www.ryzerobotics.com/ }, to represent larger robots to be used off-shore) (Figure \ref{fig:robotscombo}). This was a practical compromise because for reasons of safety, we were not permitted to use full size industrial robots and drones. The Industrial Maze module was fully open to the visitors allowing objects to be picked up and handled. The entire exhibit can be seen in Figure \ref{fig:figureEngage1} (bottom-left).

\vspace{-2mm}
\subsection{Challenges for Visitors}

Each of the modules offered a different activity and challenge for the visitors. They were of varied difficulty and offered potential for brief but engaging interaction for multiple participants simultaneously. 

1) The Wind Turbine module offered the following challenges: 
\begin{itemize}
\item One challenge was to make a Cozmo robot move to a location, retrieve a cube and return it safely to the starting point. This could be made more complex with a time limit and obstacles. This would simulate the removal of heavy tools or spare parts that a human would struggle with (Figure \ref{fig:figureEngage1} top-right).
\item If there were more than two visitors wishing to use the module, we could have them race each other to retrieve the most cubes and return them safely to their zone or they could work together to clear all of the area of cubes.
\end{itemize}

2) The Offshore Energy Platform module:
\begin{itemize}
    \item For the majority of the time, this accommodated robots displaying autonomous behavior. Visitors could engage the researchers about the behavior, but when the exhibit became busy we could have a visitor operating a Cozmo and either clearing a space for the drone to land or trying to disrupt the autonomous behavior.  The autonomous behavior would simulate the actual behavior of a full sized industrial robot on an energy platform and give the visitors a chance to see that, while autonomous robots are effective they are also not perfect and can be disrupted. Thus illustrating that there are somethings robots cannot yet handle.
\end{itemize}

3) The Industrial Maze module provided multiple activities to help illustrate some benefits of robots:
\vspace{-2mm}
\begin{itemize}
\item A visitor could operate a Cozmo and attempt to move all of the cubes from one room to the other. This task simulates hazardous waste removal.
\item A visitor could operate a Cozmo and stack the cubes in front of the Limpet sensors \cite{sayed2018limpet} thus triggering its proximity sensor and turning the LED from red to blue. This would simulate a robot being used to extinguish a fire.
\item Multiple Cozmos could be used to carry Limpets and use their light sensor to detect which cubes are illuminated. Then the other Cozmo can remove these cubes. This would simulate robots inspecting and removing hazardous material.
\item Visitors could operate an EMAT sensor (Electro Magnetic Acoustic Transducer) \cite{Duran2002emat}, they would move the sensor over a steel plate and would, by looking at a screen, try to detect damage and imperfections on the other side of the plate that was not visible, thus simulating the inspection required on offshore platforms for corrosion and damage.
\end{itemize}
\vspace{-2mm}
\subsection{Engagement Messages and Enthusiasm}

To conduct a successful engagement activity, the exhibitors need to show enthusiasm in participation. We sought volunteers from within our institutions, wishing to engage with the public and describe their work. Our team was diverse in nationality and gender.

 We trained in engagement techniques as a group and developed three core messages to convey to visitors: \textbf{1) `Let a robot do the dangerous work', 2) `Robots and renewable energy are the future' and 3) `Engineering is a diverse profession'. }

 We devised duty rotas, which carefully avoided long spells of duty over the seven-day exhibition so that the enthusiasm of the team could be frequently refreshed. 
\vspace{-2mm}
\section{Study Aims and Constraints}
\vspace{-2mm}
\subsection{Aims}
We constructed four research questions to focus our study:
 
\begin{itemize}
 \item \textbf{RQ1 Knowledge:} Does exposure to the outreach engagement exhibit impact people's knowledge about robots operating in hazardous environments? 
  \item \textbf{RQ2 Safety:} What are the visitors' perceptions in terms of rating the amount of danger for a person performing certain tasks in hazardous environments, and is this perception affected by exposure to the exhibit?
  \item \textbf{RQ3 Future Jobs:} What are the visitors' opinions in terms of rating the likelihood of a human and a robot working in hazardous environments in the future, and is this affected by exposure to the exhibit?
  \item \textbf{RQ4 Demographics Factors:} Do age group and gender have an effect on responses?

\end{itemize}
 \vspace{-2mm}
\subsection{Study Design Constraints}
Our hosts, The Royal Society, asked that collecting demographic data be kept to a minimum. A major focus of the event was on high school students (14 to 18 yrs.) soon to make decisions on future careers. However, there were particular exhibition times for families and also evenings just for adults. This age diversity, coupled with the expected high volumes, offered the opportunity to segment our participants by age group. One of the aims was promoting STEM careers. Due to the under-representation of females in some of these disciplines \cite{IET2019}, exposing any gender differences in responses might be useful to inform future interventions and engagement activities. Thus, collecting age and gender data allowed us to compare groups. As we were not collecting identifying information, age and gender, although personal data, would be non-identifiable and un-linked data. Nevertheless, we took steps in the questionnaire design to make even the collection of these two pieces of demographic data seem non-threatening for any participant. We describe these later. Also due to the expected high volume of visitors each individual engagement would be short. Any participation in our study would need to be similarly short to be proportionate. 
\vspace{-2mm}
\vspace{-1mm}
\subsection{Ethical Constraints}
Ethical considerations were of particular importance. Firstly, we wished to involve child participants and they require special consideration when obtaining consent \cite{Hay2006}. Secondly, the perceived burden of participation was carefully minimized. The decision to participate had to be an easy one (although participants were able to consider and return later if desired). 

Ethical best practice when involving children in research is to obtain the informed consent of their parent and then obtain the agreement (or assent) of the child \cite{Hay2006}. We defined a child as 15 or under in line with best practice in the domain of public engagement \cite{Hub2008}. To set a minimum age limit, we consulted with a school education practitioner. We took into account the questions and set the lower age limit at 7. Thus a child participant would be 7 to 15 years old. We encountered no pressure from visitors to include younger children, however, one enthusiastic six-year-old did participate. 



Although we only intended to collect age and gender demographic data, the presentation and wording of these fields in our questionnaire would need to be treated with sensitivity.

These ethical constraints were considered to protect the interests of participants and the reputations of the research community, our institutions and the Royal Society. These ethical issues had a central role in the study design and their proper consideration was key to obtaining ethical approval from our institution.
\vspace{-2mm}


\section{Study Method}
\vspace{-1mm}
\subsection{Main Elements and Rationale}
To allow the participation of large numbers in a fast-moving exhibition environment, we used a tablet-borne questionnaire. To minimize the time for obtaining informed consent, and participants' understanding of what they needed to do, we opted for an assisted questionnaire. Thus, participants were supported by a facilitator, who explained the quiz, obtained and recorded verbal consents, held the tablet and read out the questions from it. This also relieved participants of the burden of physically handling a questionnaire.

Protocols for administering the questionnaire, specifying how participants were approached, obtaining informed consent and, if necessary, child assent, were devised and approved as part of the study's ethical approval through our institution.

\begin{figure}
\centering
\includegraphics[width=0.75\columnwidth]{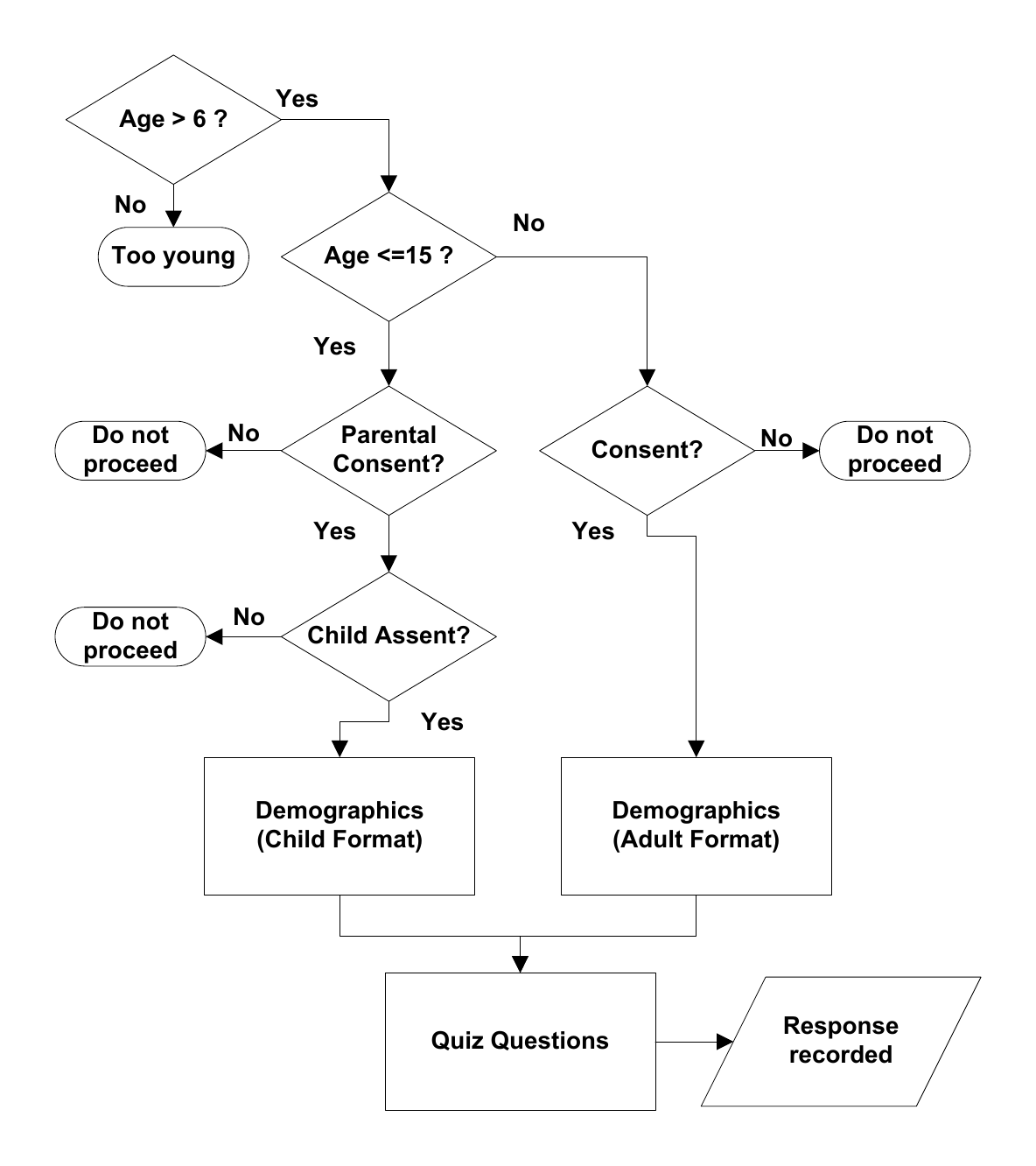}
\caption{The tablet-borne questionnaire logic flow showing decision points for the differing consent/assent requirements for child and adult participants. The details of the Child and Adult format for completing demographics fields are described in the text. The quiz questions pertained to one of four randomly picked quiz scenes Figures \ref{fig:scene-c} and \ref{fig:scenesCombo}).}~\label{fig:figureQuizWorkFlow}
\end{figure}
\vspace{-1mm}
\subsection{The Design of the Study Instrument}
\subsubsection{Quiz Format}
We decided to format the study instrument as a quiz, which participants would enjoy the challenge of answering, and view as part of their engagement with the exhibit, rather than as a survey. While this restricted us to questions suitable for a quiz, it was successful as it was rare for participants to decline. As a small reward, all participants were given an all-ages robots coloring book. 

\textit{\textbf{To address RQ1}}, we asked knowledge questions presenting scenes depicting tasks on an off-shore installation (Figure \ref{fig:scene-c}). We scored the knowledge answers using a rubric based on the answers of a panel of experts. \textbf{\textit{To address RQ2 and RQ3}}, we asked opinion questions to probe their views about the danger levels of the tasks and the future capabilities and deployment of robots to do the various tasks. Rather than attempt to exhaustively cover our research questions, we used questions to sample within those two areas.

\vspace{-1mm}
\subsubsection{Questionnaire Administration}
The tablet-borne questionnaire combined administrative and demographic questions with the main questions and followed a logic flow for consent fields based on age. We added age-based checks to cater for the consent consideration. The facilitator protocol called for consents to be obtained verbally before starting the questionnaire. The questionnaire recorded those verbal consents.
Figure \ref{fig:figureQuizWorkFlow} shows the flow involved. 
\vspace{-1mm}
\subsubsection{Sensitive Treatment of Demographic Fields}
The demographic fields were for gender and age. The issue of self-identification with gender labels can be a delicate one. We were simply interested in being able to say whether or not females were answering our quiz any differently to males. However, sex and gender are often intertwined, often unintentionally, in questionnaires \cite{Westbrook2015}. To allow our facilitators to deal sensitively with the gender field, we presented this differently for children and adults. For children, we did not confront them with the question. Instead the facilitator provided a judgment, i.e.: Question, "Participant is..." with available responses being "Male", "Female", and "Don't know". Participants 16 or over were asked to complete the field themselves without the facilitator overlooking and the item took this form: Question, "I am ..." with available responses being "Male", "Female", and "Prefer not to say". It was a deliberate design choice not to label the items as asking about gender but instead allow those answering to infer a label for the question from the response labels. In this way, using facilitator judgments, avoiding labeling the question as about gender, and providing alternative response labels, we avoided forcing any participant to self-categorize as male or female \cite{Westbrook2015}.


\begin{figure}
\centering
  \includegraphics[width=0.9\columnwidth]{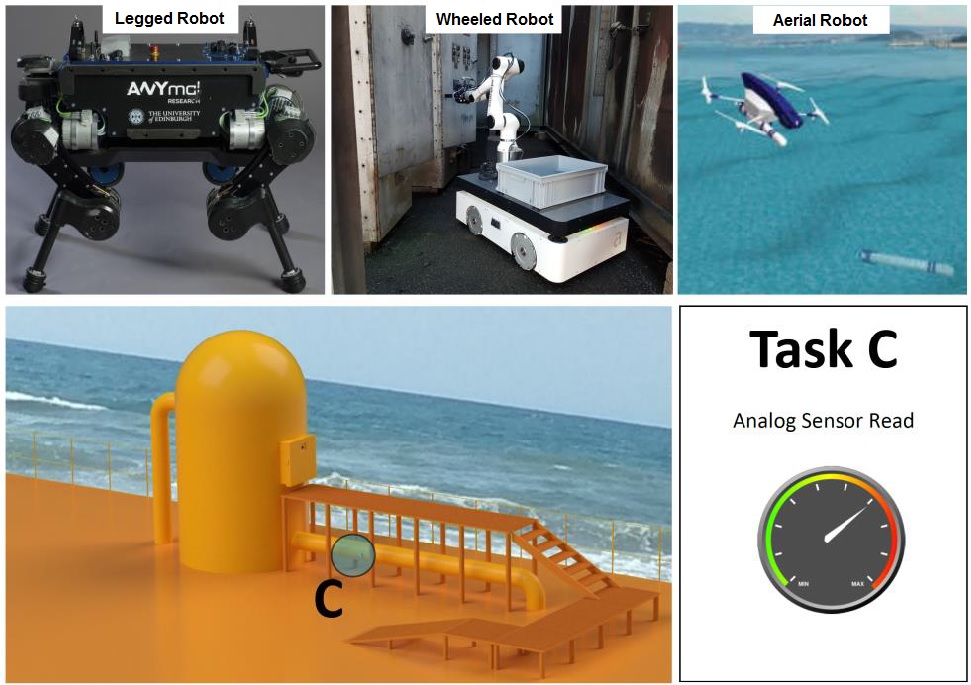}
  \caption{Example scene card held by a participant and providing the context for their answers to the quiz. Facilitators gave participants guidance on the scale of the robots and the scene. The other three scenes are described in the text and summarized in Figure \ref{fig:scenesCombo}.}~\label{fig:scene-c}
\end{figure}

\begin{figure}
\centering
  \includegraphics[width=0.85\columnwidth]{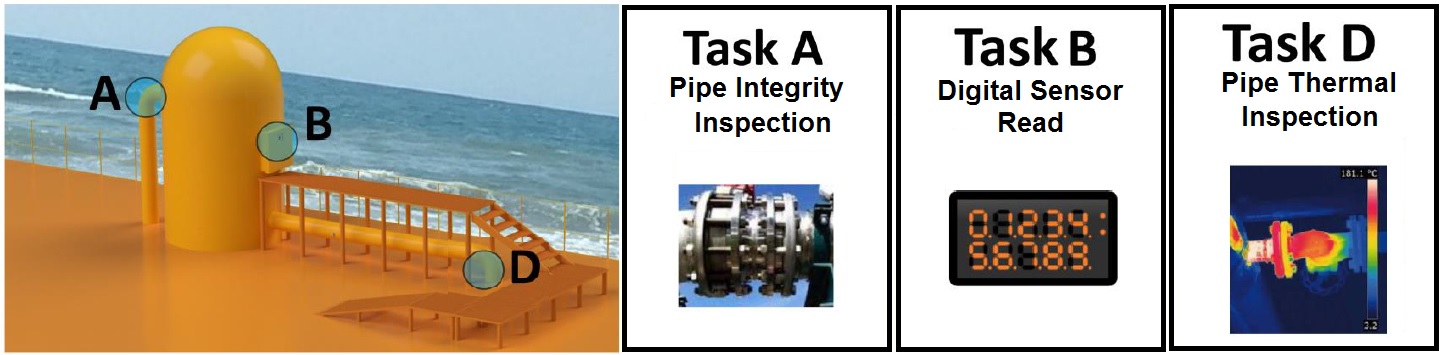}
  \caption{Scene cards A, B, and D are described by this montage containing their information pictorially summarised. See the full Scene C card in Figure \ref{fig:scene-c}. }~\label{fig:scenesCombo}
\end{figure}
\vspace{-1mm}
\subsubsection{Knowledge Items - QK1 and QK2}
To test their knowledge of robotics related to our exhibit (offshore energy asset inspection) and probe their opinions about robot capabilities now and in the future, we presented participants with one of \textit{\textbf{four question scenes}}, randomly picked, depicting four inspection scenarios as shown in Figures \ref{fig:scene-c} and \ref{fig:scenesCombo}. The four scenes were of the same difficulty and depicted a selection of inspection tasks on a block of machinery on a fictional off-shore energy platform. While probing knowledge, the questions would implicitly require participants to understand positively that only certain robots and sensors are appropriate for particular tasks, while on the other hand some are not suitable. After considering a scene, participants would all respond to the same questions. To keep participation time to a minimum, each participant was asked about only one scene (randomly selected).

The scenes and labeled images of the choice of robots were shown to participants on laminated A5 sized cards. (Figure \ref{fig:scene-c} shows an example). This allowed the scene to be in view to the participant at all times during the quiz rather than it scrolling on the tablet. Participants held the scene card. The facilitator held the tablet, read the questions and either entered the participant's answer or allowed them to choose the answer on the tablet if they wished. 

\begin{itemize}
    \item \textbf{QK1} asked them to choose from three different robots (aerial, legged, and wheeled) to carry out the task. 
    \item \textbf{QK2} asked them to choose from four different sensors for the robot to carry and to do the inspection (camera, thermometer, pressure sensor and microphone). 
\end{itemize}

\vspace{-2mm}
\subsubsection{Perception of Danger, and Opinion Items - QPD1, QOp1, and QOp2}
\begin{figure}
\centering
  \includegraphics[width=0.75\columnwidth]{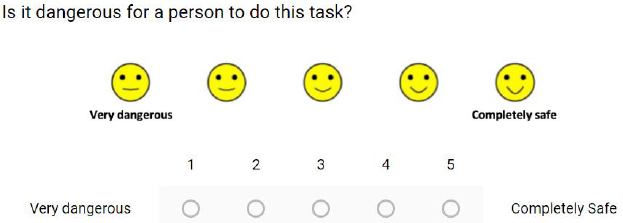}
  \caption{Example quiz question using the Smileyometer format \cite{Zaman20013smileyometer}. This shows QPD1, probing perception of danger.}~\label{fig:figure-smileyometer-question}
\end{figure}

These items used the "Smileyometer" format of Likert-style items with opposing semantic poles and icons with varying degrees of smile so as to be more user-friendly for children \cite{Zaman20013smileyometer} (Figure \ref{fig:figure-smileyometer-question}). Note how rather than the lowest rating having a sad face, it has a very slight smile because sad faces have been shown to be rarely selected by children \cite{Zaman20013smileyometer}.
To probe perception of the amount of danger that the task involved and to explore opinions about the likely adoption of robots in the future, we asked the following:
\begin{itemize}
    \item \textbf{QPD1} ``Is it dangerous for a person to do this task?'', responses: ``Very Dangerous'' (1) up to ``Completely safe'' (5). 
    \item \textbf{QOp1} ``Do you think a human will be doing this task in 10 to 15 years from now?'', responses from ``Absolutely not'' (1) up to ``Definitely yes'' (5). 
    \item \textbf{QOp2} ``Do you think a robot will be doing this task in 10 to 15 years from now?'', responses from ``Absolutely not'' (1) up to ``Definitely yes'' (5).
\end{itemize}

\vspace{-1mm}
\subsection{Experimental Design}
\subsubsection{Hypotheses}
We formed the following hypotheses:
\textit{\textbf{H1 }}- (Addressing RQ1) Participants \textit{exposed} to our exhibit will score more highly on the knowledge questions (QK1 and QK2) than those \textit{not exposed}. \textit{\textbf{H2 }}- (Addressing RQ2) Perception of danger (QPD1) will differ between the exposed and not exposed participants.  \textit{\textbf{H3}}- (Addressing RQ3) Opinions on the future of robots in hazardous environments (QOp1 and QOp2) will differ between the exposed and not exposed participants. \textit{\textbf{H4}} - (Addressing RQ4) Age group may have an effect on participant responses. \textit{\textbf{H5}} - (Addressing RQ4) Gender may have an effect on participant responses.

\begin{figure}
\centering
  \includegraphics[width=0.50\columnwidth]{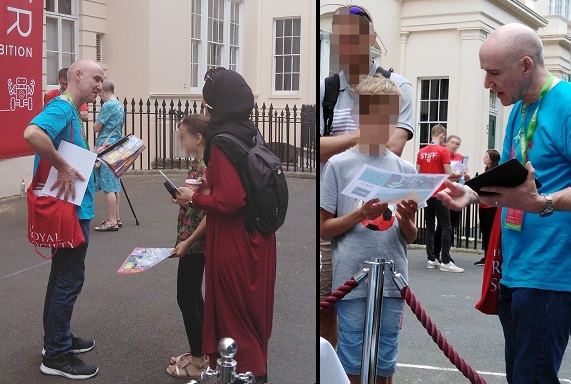}
  \caption{Child participants with their parents completing the quiz before entering the exhibition hall and engaging with the exhibit and thus part of the "\textit{Not Exposed"} group.}~\label{fig:figureOutside}
\end{figure}

Thus, \textbf{the condition} which we manipulated was \textbf{\textit{exposed} }and \textbf{\textit{ not exposed}} to engaging with our exhibit (Figure~\ref{fig:figureOutside}). The study used a between-subjects design to allow us to investigate any effect of exposure to our exhibit's engagement activities. The other factors which we explored were \textit{age group }and \textit{gender}. 

\vspace{-2mm}
\subsubsection{Age Groups}
We wished to explore variations across age groups especially as we expected to have a large number of child participants. Rather than choose age groups in an arbitrary way, we used generational boundaries. There are differing definitions of these \cite{StraussAndHowe2003, Welkey2008, Biggs2007, Pilcher1994}. We chose age groups that both fitted contemporary definitions and gave us reasonable group sizes for the data collected \cite{Bump2014}. This allowed us to keep our child participants as a single age group. We chose to combine the baby boomer generation with the pre-war generation due to the low numbers in both. \textbf{This gives age groups of 7-15, 16 - 34, 35-54, and over 54s (nominally Generation Z (GenZ), Millennials, Generation X (GenX), and Baby Boomers and the prewar generations (BabyB+)).}

\vspace{-2mm}
\subsubsection{Expert Rubric to Score Knowledge Questions}
To allow us to independently score the answers to the knowledge questions QK1 and QK2, addressing \textbf{RQ1}, we recruited seven experts in robotics and AI (6 males \& 1 female). We used convenience sampling from the academics and postdoctoral researchers at our institutions. None were authors. All were volunteers. Each completed all four questionnaires based on the four scenes. Participants' answers matching the experts were scored as 1. Where there was disagreement between the experts on the best correct answer, a weighted score was given equivalent to an answer's popularity among the expert group, e.g. for Scene B, item QK1, 3/7 experts chose "Legged" and 4/7 chose "Aerial" as the best robot for the task thus for that item the scoring rubric was: "Legged" scored 0.43; "Aerial", 0.57.  
\vspace{-1mm}

\begin{table}[]
    \centering
    \begin{tabular}{p{3.0cm}|p{1cm}|p{1cm}|p{1cm}}
        Age Group  & Exposed & Not Exposed &  Total  \\ \hline
        GenZ (7-15 yrs) & 50 & 77 & 127   \\ 
        Millennials (16-34 yrs) & 66 & 90 & 156  \\
        GenX (34-54 yrs) & 45 & 43 & 88  \\
        BabyB+ (>54 yrs)  & 27 & 25 & 52  \\ \hline
        Total & 188 &235 & 423\\ \hline
         
    \end{tabular}
    \caption{The number of participants in the two exposure conditions by Age Group. 426 quizzes were completed. Missing age values in 3 rows mean the total is 423 rather than 426.
    }
    \label{tab:sample-sizes}
\end{table}

\begin{table}[]
    \centering
    \begin{tabular}{p{1.3cm}|p{1.0cm}|p{1.3cm}|p{1.2cm}|p{1.2cm}}
        \textbf{Hypoth-esis} & \textbf{Dep. Var.} & \textbf{Indep. Var.} & \textbf{F(3,423)} & \textbf{p-value}  \\ \hline
        H1 & QK1 & Exposure & 3.152 & $.07$ \\ \hline
        H1 & QK2 & Exposure & 0.345 & $.557$ \\ \hline
        \textbf{H4} & \textbf{QK1} & \textbf{Age Gp. }& \textbf{2.882} & $\textbf{<.05}$ \\ \hline
        H4 & QK2 & Age Gp. & 1.95 & $.121$ \\ \hline
    \end{tabular}
    \caption{ Multivariate Analysis of Variance (MANOVA) with Knowledge questions QK1, QK2 as Dependent Variables (DV), and Exposure and Age Group as Independent Variables (IV). F(3,423) being the \textit{F} statistic.
    }
    \label{tab:manova1}
\end{table}

\begin{table}[]
    \centering
    \begin{tabular}{p{1.3cm}|p{1.0cm}|p{1.3cm}|p{1.2cm}|p{1.2cm}}
        \textbf{Hypoth-esis} & \textbf{Dep. Var.} & \textbf{Indep. Var.} & \textbf{F(3,423)} & \textbf{p-value}  \\ \hline
        
        \textbf{H2, H4 }& \textbf{QPD1} & \textbf{Age Gp.} & \textbf{2.866 }& $\textbf{<.05}$ \\ \hline
        \textbf{H3, H4} & \textbf{QOp1} & \textbf{Age Gp. }& \textbf{2.811} & $\textbf{<.05}$ \\ \hline
        H3, H4 & QOp2 & Age Gp. & 1.414 & $.238$ \\ \hline
        
        H2 & QPD1 & Q.Scene & 1.829 & $.141$ \\ \hline
        H3 & QOp1 & Q.Scene & 0.370 & $.775$ \\ \hline
       H3 & QOp2 & Q.Scene & 1.730 & $.160$ \\ \hline

    \end{tabular}
    \caption{ MANOVA with Perception of Danger QPD1, and Opinion QOp1 and QOp2 as DVs, and Question (Q) Scene  and Age Group as IVs. F(3,423) being the \textit{F} statistic.
    }
    \label{tab:manova2}
\end{table}

\vspace{-1mm}
\section{Results}

In this section, we report the results which we will discuss in detail in the Findings, Discussion and Future Work section.

\textbf{To test H1\footnote{H1: Participants in the \textit{exposed} condition will score more highly on the knowledge questions than those \textit{not exposed}}}, the knowledge questions QK1 and QK2 were scored according to the rubric of answer weights calculated from our expert panel's answers. We then conducted Multivariate Analysis of Variance (MANOVA) with the scores for QK1 and QK2 as Dependent Variables (DVs) and with \textit{Exposure} and \textit{Age Group} as Independent Variables (IVs). See Table \ref{tab:manova1}.  
We did not observe significant differences in participants' knowledge between the \textit{Exposed} and \textit{Not Exposed} conditions. The means (\textit{M}) and standard deviations (\textit{SD}) were as follows: QK1 Exposed, M:.63, SD:.40; QK1 Not Exposed, M:.53, SD:.41; QK2 Exposed, M:.75, SD:.33; Qk2 Not Exposed, M:.74, SD:.33. On the other hand, we saw a significant effect of Age Group on knowledge for QK1 (robot choice), but no differences were observed for QK2 (sensor choice). The posthoc test suggested that there was a significant difference between GenZ (M:.5187,SD:.4175) and BabyB+ (M:.7227,SD:.32614) ($p<.05$) for QK1.

\textbf{To initially test hypotheses H2, H3 and investigate H4}\footnote{H2: Perception of danger (QPD1) will differ between the exposed and not exposed groups.  H3: Opinions on the future of robots in hazardous environments (QOp1 and 2) will differ between the exposed and not exposed groups. H4: Age group may have an effect on participant responses.}, we conducted a MANOVA with QPD1 (Is it dangerous for a person to do this task?), QOp1 (Do you think a human will be doing this task in 10 to 15 years from now?), and QOp2 (Do you think a robot will be doing this task in 10 to 15 years from now?) as DVs and Question Scene and Age Group as IVs. Question Scene was included to investigate if that was having an effect. We use analysis of variance on our opinion rating data as has been done in other HCI studies \cite{Jun2017parametric,August2018parametric} and which is validated in Norman \cite{Norman2010parametric}. (See also Pell \cite{Pell2005parametric} and Jamieson \cite{Jamieson2004parametric} for further detail on this.) 
See Table \ref{tab:manova2}. 

For the significant results in the MANOVA, we conducted posthoc tests with Bonferroni correction. For QPD1, we found a significant difference between Millennial and GenX ratings ($p<.05$) with Millennials (M:2.77, SD:.908)  rating it more dangerous for a person to do a task as compared to GenX (M:2.40,SD:.953). For QOp1, we also found a significant difference between GenZ and GenX ($p<.05$) with GenZ (M:2.64,SD:1.07)  rating humans performing the given task in 10 to 15 years less likely than GenX (M:2.40,SD:.953).

\textbf{To further investigate H4}, we did tests as follows. The difference among the groups of generations motivated us further to investigate the differences in perception and opinions for each of the Scenes A, B, C, and D. Hence, we individually conducted MANOVAs with QPD1, QOp1, and QOp2 as DV and Generation group as IV for each of the four scenes. For reasons of space, we report here only those significant at the .05 probability threshold along with post-hoc test results: Firstly, for Scene B, QPD1 (F(3,89)=3.068, $p<.05$), with posthoc tests showing that GenZ (M:3.21,SD:1.19) rated the danger of the task significantly higher ($p<.05$) than GenX (M=2.31,SD:.946). Lastly, for Scene D, we observe significant difference for QPD1 (F(3,112)=3.274, $p<.05$), with post-hoc tests showing that Millennials (M=2.71,SD:.879) rated the danger of the task significantly higher ($p<.05$) than GenZ (M=2.09,SD:.753). 

\textbf{To further investigate H2, H3, and H4} we did the following tests. To investigate further the differences in perception and opinions between the \textit{exposed} and \textit{not exposed} conditions, we conducted further MANOVAs with QPD1, QOp1, and QOp2 as DVs and Generation and Exposure as IVs separately for Scenes A, B, C, and D.  As above, for reasons of space, we report here only those significant at the .05 probability threshold along with posthoc test results: For Scene C, we observed a significant interaction effect between Age Group and Exposure for QOp2 (F(3,109)=4.904, $p<.05$). The MANOVA showed that, interestingly, all generation groups except BabyB+ ratings of humans doing the task in Scene C were significantly lower ($p<.05$) in the \textit{Not Exposed} condition than the \textit{Exposed} condition (with BabyB+, ratings were significantly ($p<.05$) higher in the \textit{Exposed} condition). The mean and SD for BabyB+ \textit{Not Exposed} and \textit{Exposed} were M:1.33,SD:.50, \& M:3.20,SD:1.30 respectively.

\textbf{To investigate H5}\footnote{H5: Gender may have an effect on participant responses.}, we tested for significant differences between how males and females answered all 5 questions, but found none (we carried out MANOVA with DVs as the five questions and IV as Gender.) Of the 426 participants 235 were Male, 189 were Female and 2 preferred not to say (too few to analyze). The mean and standard deviation for the age of males and females were M:29.19, SD:19.54 \& M:29.01, SD:17.09 respectively. 
\vspace{-2mm}

\section{Limitations}
Firstly, our focus was robotics in hazardous environments as opposed to RAI in general. We hope in future to apply our engagement strategy to robotics in more social settings. Aside from that, the limitations in our study are chiefly those of restrictions imposed by the need to make participation lightweight and quick for participants. Using an assisted questionnaire may have introduced some facilitator bias. It is possible that unconscious bias may have caused some uneven coverage by age by facilitators. There were seven facilitators across all the days of the exhibition. Any possible bias was minimized by each receiving the same training and following the same protocol. If any bias did exist we expect that it would be largely randomly distributed across the participant groups and thus simply contribute to experimental noise rather than be systematic. Our use of four question scenes to match our exhibit's wide scope of robots and sensors may have been a confounding factor in that we had less statistical power. We discuss this particular limitation and improvements we would make to our study design below. 
\vspace{-2mm}

\section{Findings, Discussion and Future Work}
In this section, we set out specific findings and for each discuss the results leading to them. We discuss the effectiveness of our engagement strategy and the more general insights into public perception of the future of robotics and AI in hazardous environments. Where we mention generational differences, we do not claim such differences are necessarily based on social generations, they may simply be age group differences. 

\textbf{RQ1 - Knowledge: No significant effect on knowledge from exposure to the exhibit.}
Our comparison of the quiz knowledge question scores between participants \textit{exposed} and \textit{not exposed} to our exhibit show that there was no significant difference. Therefore, our hypothesis \textit{H1} was not accepted. We interpret this as being probably due to a mismatch between the knowledge that our exhibit provided to our participants and that which we probed in the two knowledge questions. Not every participant interacted over the full range of activities. Thus, in future iterations of our engagement strategy, we plan to both a) focus specific knowledge in our key engagement messages and b) ensure that any knowledge we probe in an accompanying quiz matches the key engagement messages. Additionally, to simplify analysis
and improve statistical power, rather than attempting to probe such a wide range of knowledge with four question scenes, we will adopt a sampling approach choosing one representative scene. We had expected a more even coverage in the age groups and higher overall numbers, therefore this limitation only surfaced in the light of our experience in this study.

\textbf{RQ2 - Safety: Millennials may appreciate more the benefits of having robots undertake dangerous work than GenX do.}
In QPD1, Millennials (16 to 34 yrs.) rated the tasks as more dangerous for humans than did GenX (35 to 54 yrs.).

\textbf{RQ2 - Safety: Children rated one scene more dangerous than GenX did.}
When we examined by question scene, we found in Scene B (an inspection located on an unguarded walkway), QPD1, that GenZ rated the danger of the task significantly higher than GenX did. This could be due to GenZ's lack of experience, but it might presage a future increased imperative to have robots take over more of the hazardous work.

\textbf{RQ3 - Future Jobs: Older peoples' opinions about humans doing hazardous work were not affected by our exhibit where-as opinions of younger people were.}
Our hypothesis, \textit{H3} was partially accepted as we did expose one effect of our exhibit on participants' opinions. For one question scene, C, for all generations except BabyB+, the ratings of humans doing the task (QOp1) were lower for those exposed to the exhibit than for those not exposed. i.e. they thought that it was less likely that humans would be doing the task in 10 to 15 years' time. Scene C showed an "Analog sensor read". It could well be that most of our participants felt that it was likely this fairly simple task would be within the capability of robots soon and so humans would not be doing it, whereas the other scenes showed tasks that were viewed as less straight-forward by our participants. The fact that, for the other scenes, opinions were not significantly affected by exposure could mean we lacked enough data. However, we could interpret this as either a) our interactive exhibit strategy of presenting a balanced view, aimed at visitors seeking new information activating the Scientific Literacy Model \cite{bauer1993mapping}, was not effective as it only had a partial effect on opinions, or b) experiencing our exhibit mostly confirmed pre-existing opinions about the future of RAI in hazardous environments.

\textbf{RQ3 - Future Jobs: Children were more optimistic that robots will do the dangerous work in future}.
We did discover interesting age group differences in public opinions about the future use of AI and robotics in extreme environments. In QOp1, GenX were more certain that humans would still be doing all the tasks 10 to 15 years from now than GenZ (6 to 15 yrs.). We interpret this as our young participants, either due to their youth or their knowledge, being more optimistic that humans will be able to leave more dangerous work to robots in the future. 

\textbf{RQ4 Demographic Factors: No difference in knowledge between males and females}.
In relation to our aim of using our findings to inform future engagement in terms of encouraging females into STEM careers, we found no differences between how males and females answered all five quiz questions. This can be used to reinforce our engagement messages to encourage more females to study STEM subjects as we can point out that, as far as our data in this study can show, the male and female public's knowledge in this area does not seem to differ significantly.

\textbf{Our engagement strategy did strike a chord with the visitors.} This is evidenced by comments collected by the host institution, e.g. ``\textit{Best thing was how interactive their exhibit was - for all age ranges}''; ``\textit{Many interactive stalls, lots of robots the visitors could play with so they handle the crowds well}''; and ``\textit{The robot demonstration was very hands-on, you get to try it out for yourself and see how they could work in real-life scenarios}''.
\vspace{-2mm}

\section{Conclusion}
In this paper, we described methods used in the first iteration of a high throughput interactive Robotics and AI (RAI) public engagement strategy. We designed an engaging light touch survey instrument integrated with our engagement activity to probe public knowledge and opinions about the future of RAI in extreme environments. Our results showed that exposure to our exhibit did not significantly change visitors' immediate knowledge and only slightly moved opinion. However, they did highlight some generational differences in visitors' opinions of the future development of RAI in hazardous environments. 
We aim to apply lessons learned in future iterations of our high volume RAI engagement methods. We hope this paper will encourage adoption of our approach by other groups who wish to spread accurate knowledge of RAI to the public and help to moderate some of the misinformed views of this domain that currently distort public opinion. 
\vspace{-2mm}

\begin{acks}
We thank the volunteer experts who contributed to the quiz rubric, The Royal Society,  Maurice Fallon and his team from Oxford, Milad Ramezani for photography, ORCA Hub staff and academics, and funder, EPSRC ORCA Hub EP/R026173/1. Special thanks are due to the hundreds of cheerful participants and visitors whose enthusiasm fueled us over seven days of intense public engagement and fun.

\end{acks}

\bibliographystyle{ACM-Reference-Format}
\bibliography{main}

\appendix

\end{document}